\documentclass[journal]{IEEEtran}
\ifCLASSINFOpdf
    \usepackage[pdftex]{graphicx}
    \graphicspath{{./figures/}}
\else
\fi
\ifCLASSOPTIONcompsoc
  \usepackage[caption=false,font=normalsize,labelfont=sf,textfont=sf]{subfig}
\else
  \usepackage[caption=false,font=footnotesize]{subfig}
\fi
\usepackage{mathtools}

\usepackage{algorithm}
\usepackage{algorithmic}
\usepackage{xcolor}
\usepackage{xspace} 

\newcommand{\cut}[1]{}

\newcommand{\sys}{MeNTT\xspace}

\newsavebox{\ieeealgbox}

\begin{document}
%
\title{\sys: A Compact and Efficient Processing-in-Memory Number Theoretic Transform (NTT) Accelerator}
%
%
%

\author{Dai~Li$^*$,~\IEEEmembership{Student Member,~IEEE,}
        Akhil~Pakala$^*$,~\IEEEmembership{Student Member,~IEEE,}
        and~Kaiyuan~Yang,~\IEEEmembership{Member,~IEEE}
\thanks{$^*$These authors contributed equally.} 

\thanks{The authors are with the Department
of Electrical and Computer Engineering, Rice University, Houston,
TX, 77005, USA. email: \{dl37, arp5, kyang\}@rice.edu. (Corresponding author: Kaiyuan Yang)}
}

%
%

\markboth{Transactions on Very Large Scale Integration (VLSI) Systems}%
{Shell \MakeLowercase{\textit{et al.}}: Bare Demo of IEEEtran.cls for IEEE Journals}
%



\maketitle

\begin{abstract}
Lattice-based cryptography (LBC) exploiting Learning with Errors (LWE) problems is a promising candidate for post-quantum cryptography. Number theoretic transform (NTT) is the latency- and energy- dominant process in the computation of LWE problems. This paper presents a compact and efficient in-MEmory NTT accelerator, named \sys, which explores optimized computation in and near a 6T SRAM array. Specifically-designed peripherals enable fast and efficient modular operations. Moreover, a novel mapping strategy reduces the data flow between NTT stages into a unique pattern, which greatly simplifies the routing among processing units (i.e., SRAM column in this work), reducing energy and area overheads. The accelerator achieves significant latency and energy reductions over prior arts.
\end{abstract}


%
\IEEEpeerreviewmaketitle

\section{Introduction}
Most contemporary public-key cryptographic primitives like RSA and Elliptic Curve Cryptography (ECC) rely on the difficulty to solve integer factorization and discrete algorithms. However, with the advent of quantum computers, these problems are expected to be solvable, such as using Shor's algorithm~\cite{shor_polynomial-time_1996} in polynomial time. Therefore cryptographic algorithms that are resistant to potential attacks using quantum computers are being studied by researchers around the globe. Lattice-based Cryptography (LBC) has emerged as abprime candidate among the Post-quantum protocols.

\begin{figure}[t]
      \centering
      \includegraphics[trim={0 0cm 0 1cm},clip,width=8.2cm]{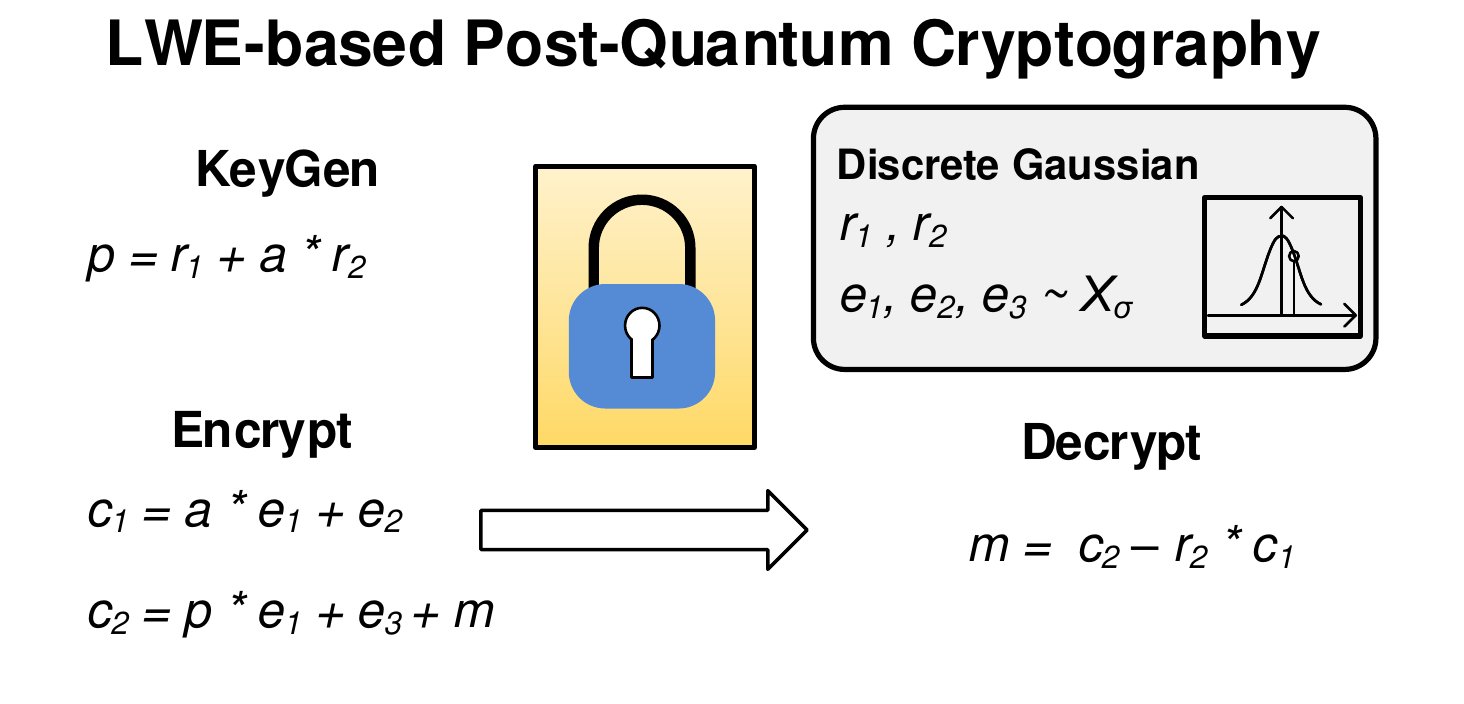}
      \vspace{-3ex}
      \caption{The basic scheme of lattice-based cryptography.}
      \label{F_concept}
   \end{figure}

Lattice-based protocols have come to light because of the hardness of inherent Learning with Errors (LWE) problems~\cite{regev_lattices_nodate,lyubashevsky_ideal_2010}. Module LWE and Ring LWE are two primary variants of the LWE problem. Ring LWE secures the message using polynomial operations between secret key and public key along with error addition (Fig.~\ref{F_concept}). Polynomial multiplication is performed using Number theoretic transform (NTT)~\cite{cooley_algorithm_1965}, an FFT like structure except that the operations performed are modular arithmetic. In a typical Homomorphic/PQC system, sampling and NTT are the two main operations in Ring LWE schemes and they take a similar amount of time~\cite{mert_design_2020}. 
Recent studies have significantly improved the energy and area efficiency of random samplers, e.g., our recent work MePLER~\cite{MePLER} achieved 20.6-pJ per sampler energy efficiency and 85.9-MSample/s constant throughput in custom 65-nm hardware. This paper thus focuses on NTT acceleration in Ring LWE 
With n being power of 2, time complexity for computing NTT is $O(n\log{n})$. Theoretically, n/2 modulo multiplications can be performed in parallel for each NTT. However, it requires accessing elements in parallel, which proves to be the bottleneck. Most of the existing NTT designs in FPGA~\cite{poppelmann_area_2014} or digital ASIC have been optimizing the dataflow to enhance the overall performance and efficiency for a certain configuration of butterfly units and memory banks. The hardware configuration underlines a fundamental trade-off between area and efficiency. For example, \cite{banerjee_23_2019,roy_compact_2013} minimize the area overhead by sequentially accessing and sharing a butterfly unit, while \cite{song_leia_2018} attempt to layout and operate several butterfly units in parallel. Both approaches could not completely address the issue of designing a compact and parallel accelerator. 

Processing-in-Memory (PIM)~\cite{verma_-memory_2019} is an emerging technology for memory-constrained computation, owing to its capabilities of highly parallel computing with amortized energy for memory accesses and logic operations. PIM also enables enhanced data locality, avoiding frequent data transfers to and from the computing units. These additional capabilities, while retaining the compact nature of the memory, make it a promising technology for accelerating NTT~\cite{nejatollahi2020cryptopim}.

While PIM accelerators based on beyond-CMOS memory devices, such as ReRAM or MRAM, hold great promise for future memory-centric computing with superb density and non-volatility. SRAM-based PIM accelerators solely based on mainstream CMOS technologies undoubtedly represent a clear path towards reliable mass productions and robust operations. SRAM PIM allows low-voltage read, write, and logic operations for energy savings, and could always take advantage of the latest CMOS processes. Further, the larger footprint of SRAMs is amortized by the relatively large PIM peripherals in practical implementations. 

In this paper, we present \sys, an in-6T-SRAM NTT/INTT accelerator with boosted area and energy efficiency. Our key contributions include:
\begin{itemize}
  \item We develop a novel protocol to perform bit-serial modular addition\slash  subtraction\slash multiplication arithmetic in 6T SRAM array with massive parallelism and less steps.
  \item We propose a mapping strategy to optimize the dataflow between NTT and INTT stages and dramatically reduce the routing overhead for large NTT/INTT.
  \item The MeNTT accelerator is rigorously evaluated in TSMC 65nm LP process, through a combination of transistor-level post layout simulation for the memory and post-layout for digital logic by Design compiler

\end{itemize}

The rest of the paper is organized as follows. Section II provides the necessary background for Ring LWE and NTT. Section III discusses the implementation of modular arithmetic in 6T SRAM, the data flow between the NTT/INTT stages and routing technique. Section IV shows the evaluation results and comparison with prior arts. Section V concludes the paper.

\section{Background and Related Work}
This section covers mathematical background of the Ring LWE problem and the NTT/INTT algorithm based polynomial multiplication. It also provides a brief overview on recent PIM research, especially those for cryptographic acceleration.

  \begin{figure}[t]
      \centering
      \includegraphics[width=0.98\columnwidth]{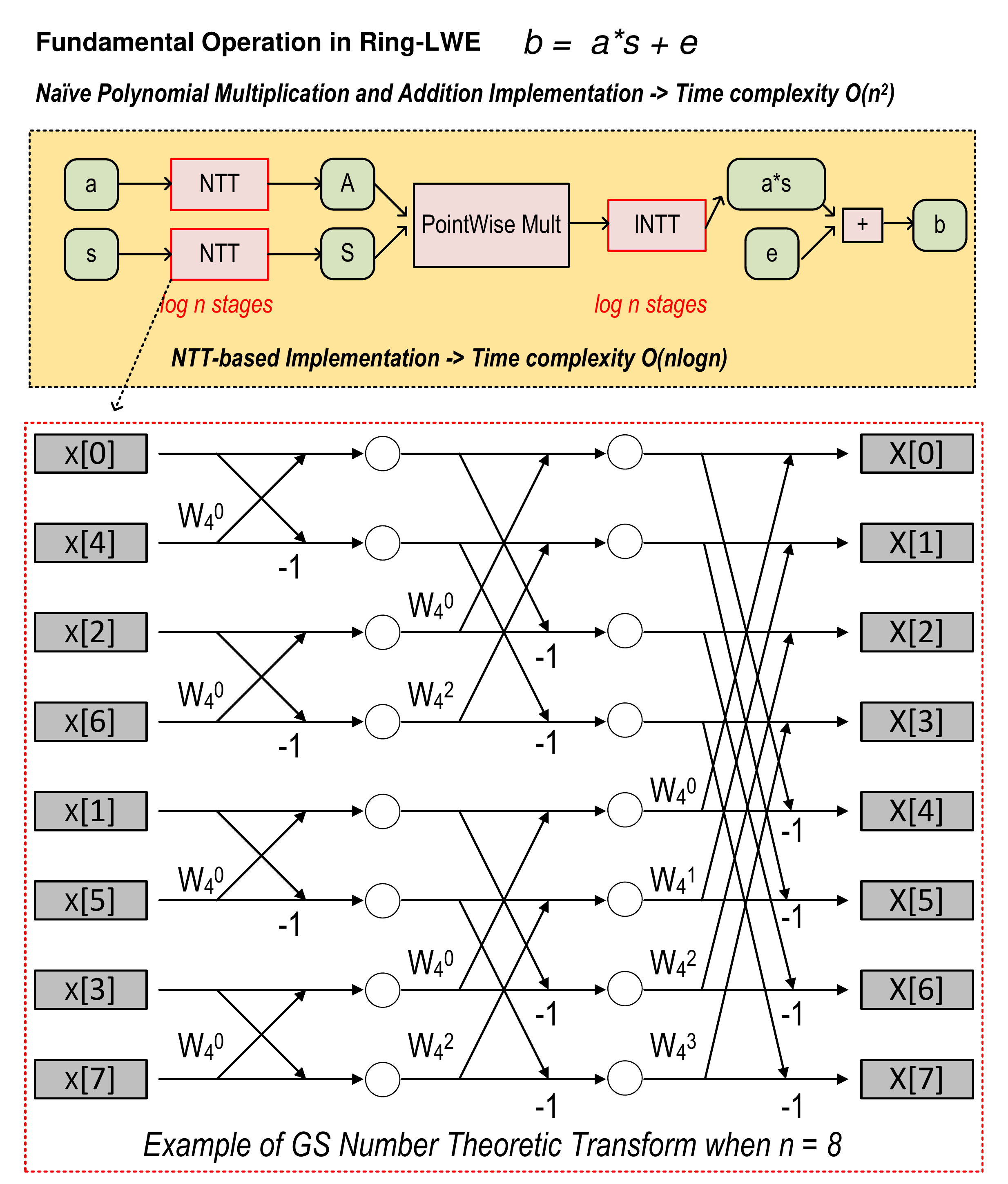}
      \vspace{-2ex}
      \caption{The overall flow of Ring-LWE and polynomial multiplication (NTT).}
      \label{F_flow}
   \end{figure}
   
\subsection{Ring Learning with Errors (LWE)}
Let the pair of vectors \textbf{(a,b)} be related by the equation ${b=a*s+e}$, where \textbf{a} is a randomly sampled vector from $R_{q}$, \textbf{e} is a error vector sampled from a Gaussian distribution. Here $R_{q}=Z_{q}[x]/(x^{n}+1)$ is a ring of polynomial where n is a power of 2~\cite{buchmann_high-performance_2016}, q is a prime number. Ring LWE~\cite{lyubashevsky2012lattice} states that it is difficult to find secret vector $\textbf{s} \in R_{q}$ given the pair (\textbf{a,b}). Here, a*s is a polynomial multiplication, and is performed by transforming both a and b in NTT domain (Fig.~\ref{F_flow}).

\subsection{Number Theoretic Transform (NTT)} 
Let a,s are two polynomials sampled from $R_{q}$, whose coefficients are in range [0,q) where q is prime number. We denote NTT coefficients of polynomial $(a=a_{n-1}x^{n-1}+a_{n-2}x^{n-2}+...+a_{0})$ as $\hat{a}_{n-1},\hat{a}_{n-2},..,\hat{a}_{0}$ respectively.  
 \begin{equation}
  \centering b = INTT((NTT(a)*NTT(s))
  \label{eq:Poly_mult}
 \end{equation}
 
Polynomial multiplication of b=a*s is done using eq \ref{eq:Poly_mult}.After transforming both vectors a and s in NTT domain, NTT coefficients of b are calculated by coefficient wise multiplication of a and s. Original coefficients are calculated back by transforming $\hat{b}$ using Inverse NTT.
 \begin{equation}
  \centering \hat{b} = \sum_{i=0}^{N-1} (\hat{a}_{i}*\hat{s}_{i})x^{i}
  \label{tab:coeff_wise_mult}
 \end{equation}
 
There are two popular variants of butterfly optimizations for the acceleration of polynomial multiplication: Cooley-Tukey~\cite{roy_compact_2014} and Gentleman-Sande. The former is adopted in MeNTT and described in Algorithm \ref{alg:NTT}. INTT computation is similar to NTT except that the twiddle factors are $w^{-n}$ instead of $w^{n}$. 

 \begin{algorithm}[t]
\begin{algorithmic}[1]
\REQUIRE Given polynomial $a \in R_{q}$, n-th root of unity $w_{n}$ in $ Z_{q}$
\ENSURE Polynomial $\hat{a} = NTT(a)$ such that $\hat{a} \in R_{q}$
\STATE $\hat{a} \leftarrow PolyBitRev(a) $
\FOR{(stage =1; stage$\leq \log_2n$; stage = stage+1)}
\STATE $m \leftarrow 2^{stage}$
\STATE $w_{m} \leftarrow w_{n}^{n/m}$
\FOR{(k =0; $k<n$; k=k+m)}
\STATE $w \leftarrow 1$
\FOR{(j =0; $j<m/2$;j=j+1)}
\STATE $t \leftarrow w.\hat{a}[k+j+m/2]\;mod\;q$
\STATE $u \leftarrow \hat{a}[k+j]$
\STATE $\hat{a}[k+j] \leftarrow u+t\;mod\;q$
\STATE $\hat{a}[k+j+m/2] \leftarrow u-t\;mod\;q$
\STATE $w \leftarrow w.w_{m}\;mod\;q$
\ENDFOR
\ENDFOR
\ENDFOR
\RETURN $\hat{a}$

\end{algorithmic}
\caption{ NTT Multiplication with Cooley-Tukey Method}\label{alg:NTT}
\end{algorithm}
 
 \subsection{Processing in Memory for Cryptographic Acceleration}

Like many other memory-centric computation problems, such as deep learning, NTT computing faces the ``memory walls" because of the energy and throughput bottleneck between logic and memory~\cite{horowitz_1.1_2014}. To alleviate this problem, a new computing paradigm with in- and near-memory computing emerged to reduce data movement, amortize memory access energy, and energy-efficient mixed-signal logic operation within a memory array~\cite{verma_-memory_2019}.

\begin{figure}[t]
      \centering
      \includegraphics[trim={0 0 0 0.6cm},clip,width=0.75\columnwidth]{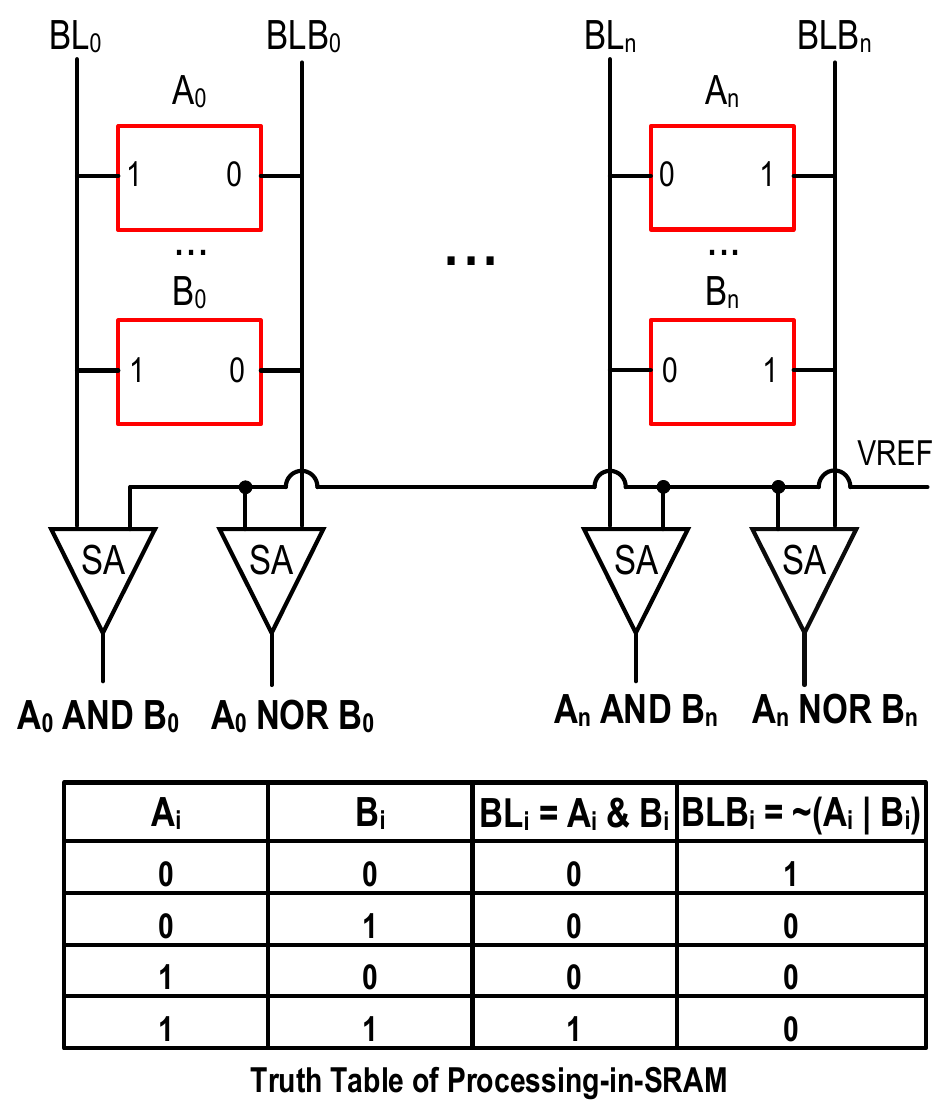}
      \vskip -2ex
      \caption{Basic bit-wise logic operations in 6T SRAM by accessing two rows simultaneously~\cite{jeloka_28_2016}.}
      \label{F_sram}
\vskip -2ex
   \end{figure}

In cryptographic computing, the requirement on PIM is different from that of machine learning applications because of its zero tolerance to compute errors. Thus, bit-parallel and bit-serial operations are more suitable than the lossy computing mechanisms in current, charge, or voltage domains~\cite{chen_cap-ram_2021,kang_energy-efficient_2014,valavi_64-tile_2019,ISAAC}.~Recent Digital in-SRAM architectures~\cite{kim_colonnade_2021,chih_164_2021} have been designed to compute with full precision and high parallelism. However, these architectures are designed to compute MAC operations and are not suitable to output multiple modular arithmetic in parallel. Bit-serial in-SRAM logic performed by accessing two words simultaneously (Fig.~\ref{F_sram}) is first utilized to modern cryptography accelerators by~\cite{recryptor}, which performs very wide word bitwise logic and finite field arithmetic in memory. However, it does not provide the high parallelism expected for NTT acceleration. Performing arithmetic in bit-serial fashion was introduced in~\cite{eckert_neural_2018}, which is promising for NTT and modular arithmetic because it achieves full accuracy computing with massive parallelism, by spending more clock cycles in each arithmetic operation. \cite{nejatollahi2020cryptopim} leveraged a similar technique with projected high-performance and high-density ReRAM devices. It employs an unfolded architecture with straightforward control and routing schemes, which is possible with high-density ReRAM devices and slim peripherals, but will lead to an unacceptably large area if implemented on SRAM. On the other hand, recent ASIC NTT accelerator~\cite{song_leia_2018} has tried computing multiple butterfly operations in parallel and moving buffers closer to the computing units. However,the extensive use of registers for local buffering comes with a high area overhead. Thus, achieving the desired in-memory computing performance and efficiency gains with compact physical footprint and cost is the primary goal of this work. 

\section{Data Storage and Arithmetic Flow in \sys}
\begin{figure}[t]
      \centering
      \includegraphics[trim={0 0 0 0},clip,width=\columnwidth]{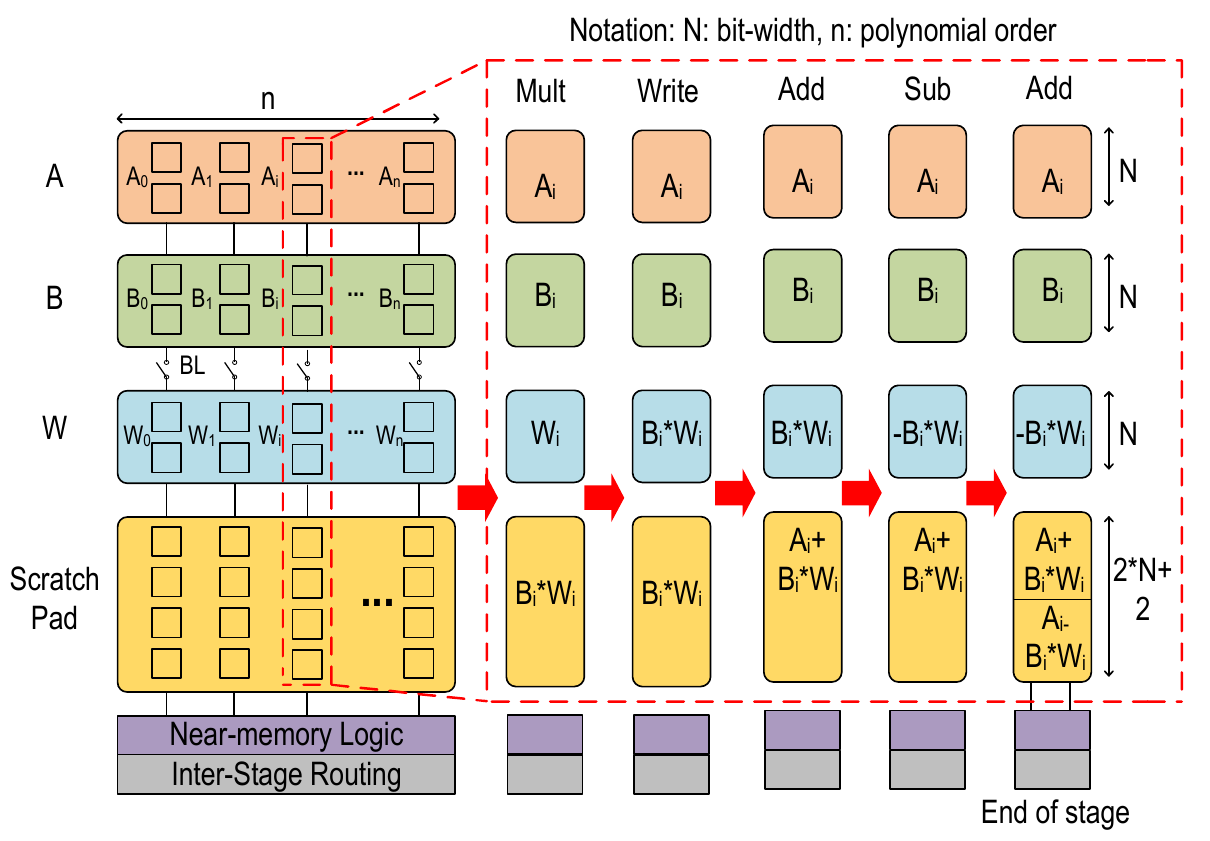}
      \vskip -2ex
      \caption{Data arrangement and operation sequence in an NTT stage.}
      \label{F_data_arr}
   \end{figure}

{\sys} performs complete polynomial multiplication computation in NTT and INTT with all the modular arithmetic steps in and near a 6T SRAM array, as shown in Fig. ~\ref{F_diagram}. The SRAM array functions both as data storage and a computing unit. Inter-column router route data at the end of one stage into corresponding columns for the computation of the next stage. SRAM peripheral and controller are specifically designed for in- and near-memory modular arithmetic. During each stage in the radix-2 NTT, each column works like a butterfly unit, thus enabling massively paralleled computing to support a large number of points. The operands and intermediate results are all stored in the same array with an allocation strategy shown in Fig.  ~\ref{F_data_arr}. As a result, the width n of the array represents the maximum number of points of NTT, while the number of rows is related to the supported bit width, as shown in Fig.~\ref{F_data_arr}. The modular addition, subtraction, and multiplication are performed in a bit-serial manner, similar to the generic bit-serial logic achieved in~\cite{eckert_neural_2018}, but with significantly reduced steps and energy-optimized for modular arithmetic. The high parallelism enabled by the bit-serial approach improves the overall performance and energy efficiency of NTT operations with a large number of points. In our implementation, a single SRAM bank with 162 by 1024 cells is designed to support 1024-point and 32-bit NTT operations. The physical layout of the wide SRAM array can be folded as shown in Fig.~\ref{F_diagram}, in order to reduce word line and inter-column routing length, and maintain a proper aspect ratio. 6T SRAM is desired because of its maturity and high density. 

Modular arithmetic differs from regular arithmetic as it pays more attention to overflow and underflow issues. Most previous works take the strategy to calculate the result in an integer field, then reduce it to the desired finite field. Barrett reduction and Montgomery reduction (Algorithm \ref{alg:barrett_mul}, Algorithm \ref{alg:montgomery_mul}) are the two most common approaches to reduce numbers into a finite field. {\sys} proposes a reduce-on-the-fly technique for modular addition, subtraction, and multiplication.

\subsection{Near-Memory Bit-serial Comparator and Reduction}

\begin{figure*}[t]
      \centering
      \includegraphics[trim={0.1cm 0 0 0},clip,width=1.8\columnwidth]{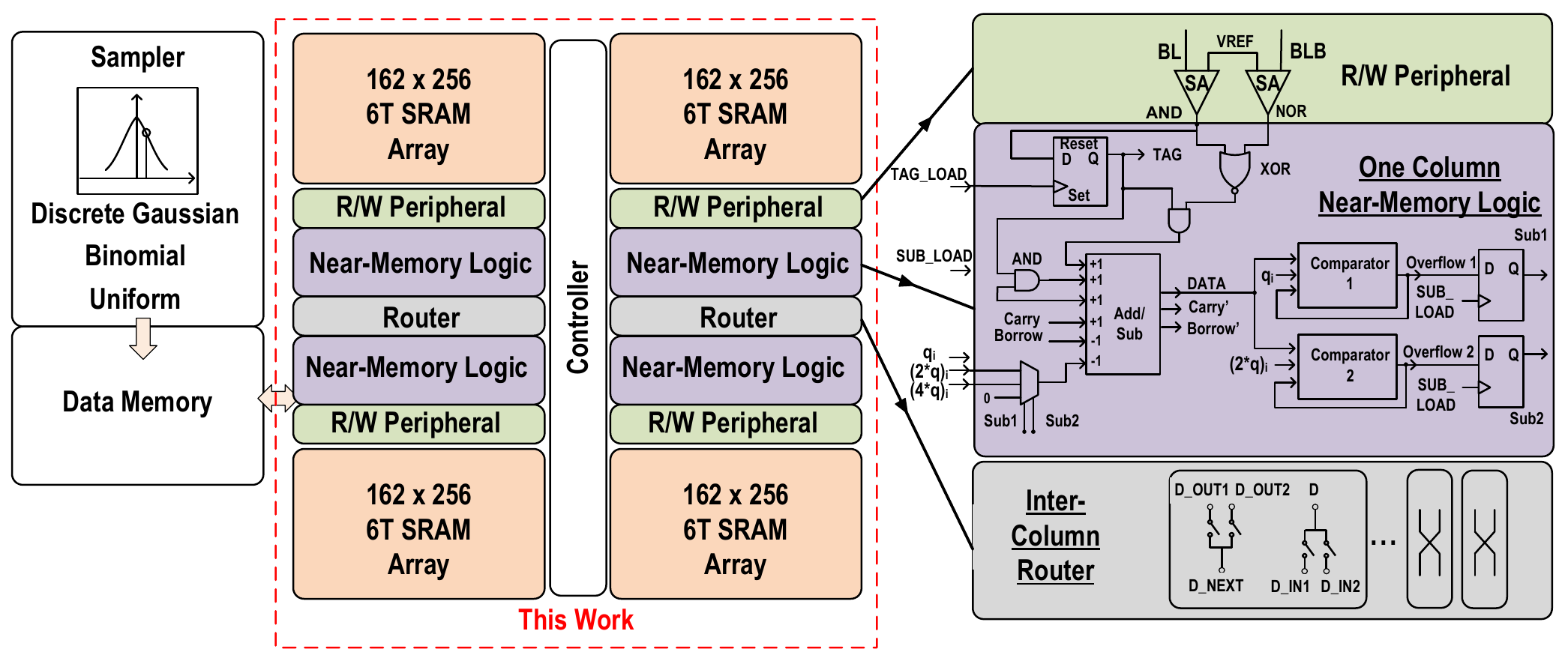}
      \vskip -2ex
      \caption{Diagrams of {\sys}  and custom circuitry for the key peripherals.}
      \label{F_diagram}
   \end{figure*}

\begin{figure}[t]
      \centering
      \includegraphics[trim={0 0 0 0},clip,width=\columnwidth]{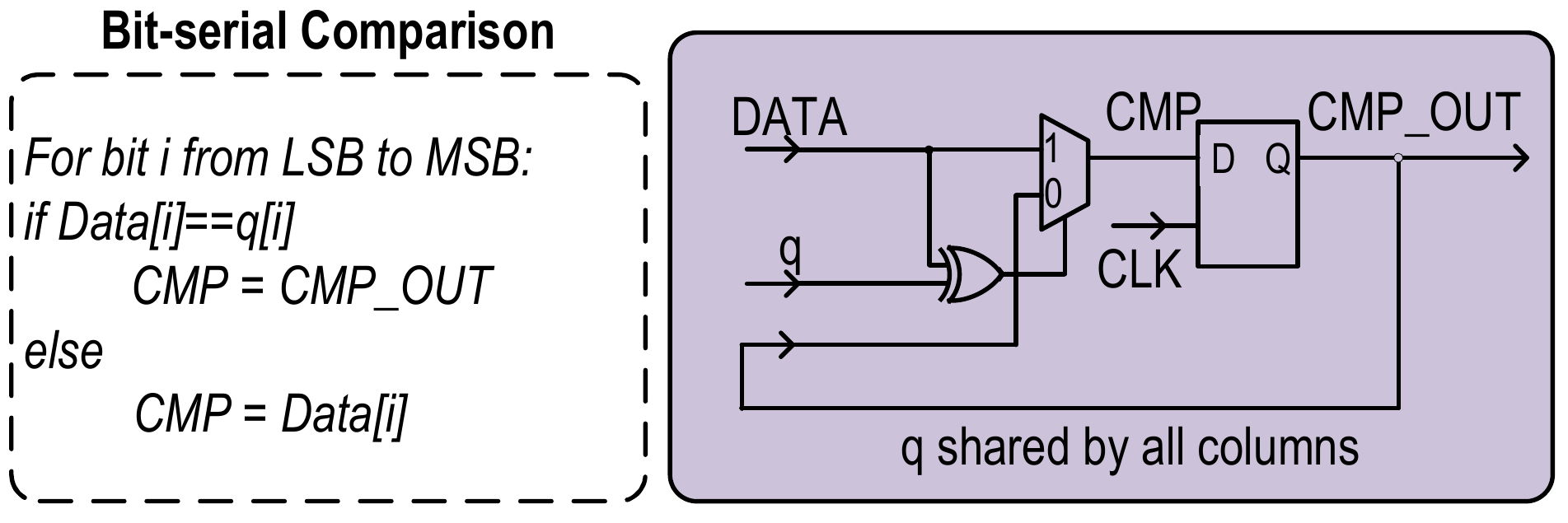}
      \vskip -2ex
      \caption{Bit-serial comparator operation in column peripheral.}
      \label{F_cmp}
   \end{figure}

We execute addition or accumulation from LSB to MSB in our bit-serial arithmetic operation. In order to keep the final result within field range, our reduction scheme utilizes the bit-serial comparator (Fig. \ref{F_cmp}) to keep track of the overflow condition of the result from the last operation. The comparison between temporary result and field limit `q' starts from LSB to MSB cycle by cycle as bit-serial addition, subtraction or multiplication takes place. Reduction is applied based on the comparison result (Fig.~\ref{F_diagram}) by applying a subtraction logic together with the adder. Only when reduction is required, the subtraction of q is enabled. For addition and subtraction, a 'raw' result is calculated initially to evaluate an overflow. Then a 'real' step is done to calculate the actual result with reduction. For multiplication, the partial sum is compared with q, while reduction is applied in the next round of accumulation with proper scaling. 

\subsection{In-Memory Modular Addition/Subtraction}

\begin{figure*}[t]
      \centering
      \includegraphics[trim={0 0cm 0 0},clip,width=14cm]{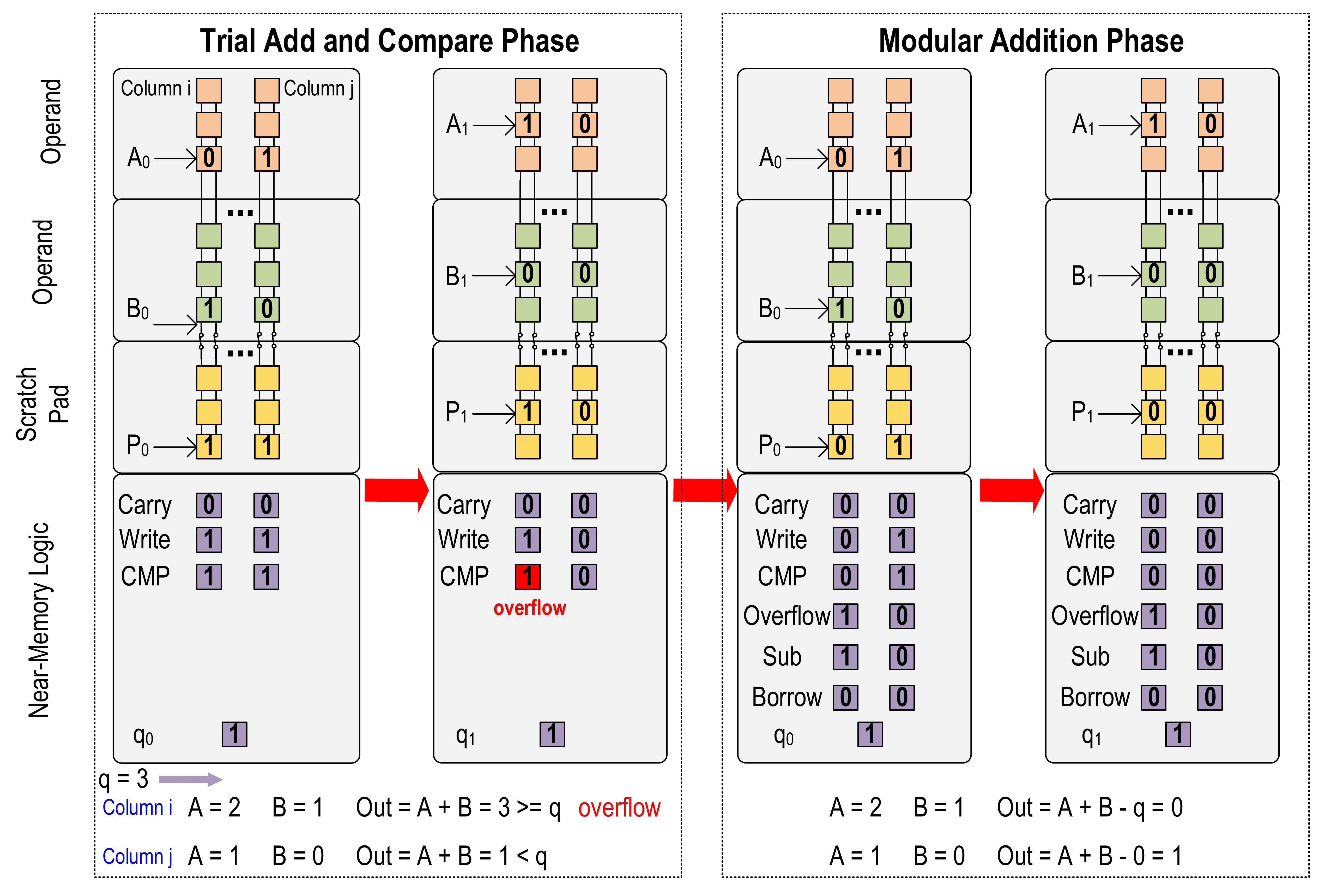}
      \vskip -2ex
      \caption{Illustration of in-memory two-bit modular addition.}
      \label{F_addsub}
   \end{figure*}

The proposed modular addition and subtraction consist of a regular round of addition/subtraction with overflow/underflow detection called "Trial Add and Compare Phase", and a round of modular reduction called "Modular Addition Phase", as depicted in (Fig. \ref{F_addsub}). The WL driver activates corresponding bits of two operands A and B sequentially. The value on BL will be A AND B, while the value on BLB will be A NOR B. These values are read using traditional sense-amplifier and given as inputs to near-memory. The peripheral maintains one bit Carry for the addition, and acts like a full adder which adds A and B up sequentially and writes back to the array. In ``Trial Add and Compare Phase", in the meantime of addition, the peripheral reads one bit of q, and makes a comparison sequentially. An overflow bit is set if the sum of A and B is larger than q. In that case, column will calculate A+B-q during "Modular Addition Phase". Otherwise, column only calculates A+B to avoid timing side-channel leakage

Modular subtraction is done similar to modular addition by transforming one of the input B into it's 2's complement first. This can be done by setting the initial carry to 1 and using the value from BLB as input. During "trial add and compare" phase after 2's complement, an underflow bit is recorded. Based on the underflow bit value, addition of q is done during second-round subtraction to keep the result value correct. Addition requires 2*(N+1) cycles while the subtraction requires an extra N+1 cycles for 2's complement calculation.

\newcommand{\floor}[1]{\left\lfloor #1 \right\rfloor}
\newcommand*\ShiftRight{\gg}

\begin{algorithm}[t]
\begin{algorithmic}[1]
\REQUIRE: $x, y \in Z_{q},\ q, m\ and\ k\ such\ that\ m = \floor{2^{k}/q}$
\ENSURE z = (x.y) mod q
\STATE $ z \leftarrow x\cdot y$
\STATE $ t \leftarrow (z\cdot m) \gg \ k$
\STATE $ z \leftarrow z - (t \cdot q)$
\IF{($z \geq q$)}
\STATE $ z \leftarrow z-q$
\ENDIF
\RETURN z
\end{algorithmic}
\caption{Barrett Modular Multiplication}\label{alg:barrett_mul}
\end{algorithm}

\begin{algorithm}[t]
\begin{algorithmic}[1]
\REQUIRE: $x, y \in Z_{q},\ q, r\ and\ k\ such\ that\ r > q,\ gcd(r,q)=1,\ k=\frac{r(r^{-1}mod\ n)-1}{n}$
\ENSURE: $c = x\cdot y \ mod\ q$
\STATE $\overline{a}=a\cdot r\ mod\ q$
\STATE $\overline{b}=b\cdot r\ mod\ q$
\STATE $ x = \overline{a}\cdot \overline{b}$
\STATE $ s = x \cdot k \ mod \ r$
\STATE $ t = x+s \cdot q$
\STATE $ u=\frac{t}{r}$
\IF{($u \geq q$)}
\STATE $ \overline{c} = u-q$
\ELSE
\STATE $ \overline{c} = u$
\ENDIF
\STATE $c=(\overline{c}\cdot r^{-1} mod \ n)$
\RETURN c
\end{algorithmic}
\caption{Montgomery Modular Multiplication}\label{alg:montgomery_mul}
\end{algorithm}

\begin{algorithm}[t]
\begin{algorithmic}[1]
\REQUIRE $x,y \in Z_{q}$
\ENSURE $z=x+y \; mod \; q$\
\STATE Trial Add and Compare Phase:
\STATE  $s \leftarrow x+y$
\STATE  $cmp \leftarrow s \geq q$
\STATE Modular Addition Phase:
\IF{($cmp = 1$)} 
\STATE $z \leftarrow x+y-q$
\ELSE
\STATE $z \leftarrow x+y$
\ENDIF
\RETURN z
\end{algorithmic}
\caption{Modular Addition}\label{alg:modular_add}
\end{algorithm}

\subsection{In-Memory Bit-Serial Modular Multiplication}
\begin{figure}[t]
      \centering
      \includegraphics[trim={0cm 0cm 0cm 0cm},clip,width=6cm]{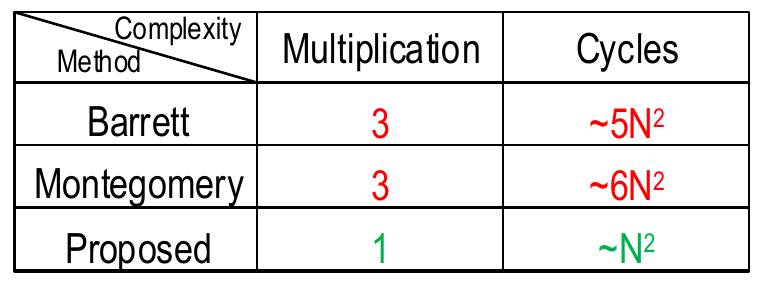}
      \vskip -1ex
      \caption{Comparison of conventional modular multipliers and the proposed one in MeNTT.}
      \label{F_modmultcost}
   \end{figure}

Modular multiplication is the most time-consuming part of the NTT. Traditionally, modular multiplication is performed by computing the raw product first, then going through a Barrett or Montgomery reduction(Fig. \ref{F_modmultcost}). In previous PIM works \cite{eckert_neural_2018}, this involves three normal multiplications and also takes extra redundant cycles as well as space. Although in some special cases \cite{nejatollahi2020cryptopim} this reduction can be simplified, there is no general optimization for the modular multiplication using traditional reduction approaches.

\begin{algorithm}[t]
\begin{algorithmic}[1]
\REQUIRE $x,y \in Z_{q}$, where $y_{k}$ represents k-th bit of N-bit $y$
\ENSURE z = (x.y) mod q
\STATE $psum=0, overflow\_4q=0, overflow\_2q=0$
\FOR{($k=N-1$; $k\geq0$; $k=k-1$)}
\FOR{($j=0;j < N; j=j+1$ )}
\IF{($overflow\_4q=1$)}
\STATE $psum_{j} \leftarrow (psum_{j}\ll1)+(x_{j}*y_{k})-q_{j-2}+carry-borrow$

\ELSIF{($overflow\_2q=1$)}
\STATE $psum_{j} \leftarrow (psum_{j}\ll1) + (x_{j}*y_{k})-q_{j-1}+carry-borrow$
\ELSE  
\STATE $psum_{j} \leftarrow (psum_{j} \ll1) + (x_{j}*y_{k})+carry-borrow$
\ENDIF
\STATE $write_{j} = LSB(psum_{j})$
\STATE $carry = (psum_{j}>1)?1:0$
\STATE $borrow = (psum_{j}<0)?1:0$
\STATE $cmp_{2q} = (psum_{j}==q_{j-i})?cmp_{2q}:psum_{j}$ 
\STATE $cmp_{4q} = (psum_{j}==q_{j-2})?cmp_{2q}:psum_{j}$
\ENDFOR
\IF{($cmp_{4q}==1$)} 
\STATE $overflow\_4q \leftarrow 1$
\ELSE
\STATE $overflow\_4q \leftarrow 0$
\ENDIF
\IF{($cmp_{2q}==1$)} 
\STATE $overflow\_2q \leftarrow 1$
\ELSE
\STATE $overflow\_2q \leftarrow 0$
\ENDIF
\ENDFOR
\IF{($psum\,\geq\,q$)} 
\STATE $psum \leftarrow psum-q$
\ENDIF
\STATE $z \leftarrow psum$
\RETURN z
\end{algorithmic}
\caption{Modular Multiplication}\label{alg:modular_mul}
\end{algorithm}

  \begin{figure}[t]
      \centering
      \includegraphics[trim={0 0cm 0 0},clip,width=\columnwidth]{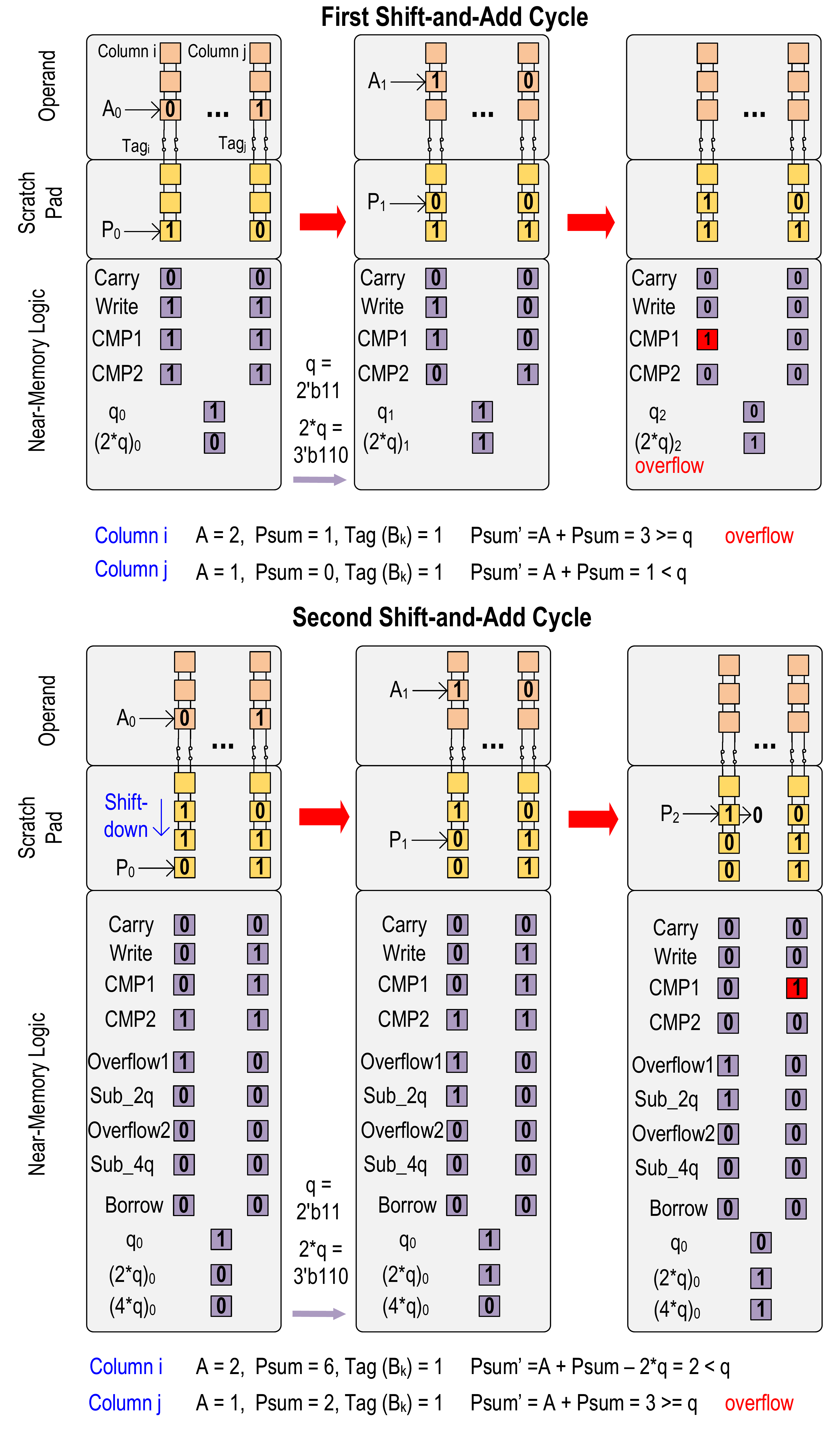}
      \vskip -2ex
      \caption{Illustration of in-memory 2-bit modular multiplication (q is 2'b10, Tag is 1 for two cycles).}
      \label{F_modmult}
   \end{figure}

This work proposes a fast bit-serial multiplication scheme to complete a modular multiplication in (N+1)\textsuperscript{2} cycles (Fig. \ref{F_modmult}). The overall algorithm is described in Algorithm 5. The multiplication is decomposed into shift and add operations. There is a Tag bit which is similar to the approach of \cite{eckert_neural_2018}. The shift is done by choosing a different address for rows in each round. The addition is controlled by the Tag bit from operand B and computed in the peripheral with carry and borrow. We take advantage of the bit-serial computation flow and compare every partial sum with q and 2*q. A reduction will be performed in the next cycle to make sure the partial sum is contained in the correct range. The reduction is enabled by the subtraction and borrow circuits, and controlled by the two overflow bits. In this way, the modular multiplication can be completed in parallel with the shift-and-add itself with very little space and time overhead.
In order for the algorithm to be valid, operand A has to be smaller than q/2. Therefore we use 2*q instead of q as the field constant. Then \sys \ reduce the partial sum below q with an extra cycle. A switch is controlled by the Tag to separate the operand A from influencing the BL readout result. It decides whether the partial sum is added by an operand or kept the same. The reduction with q or 2q is done in peripheral with simple digital logic shown in Fig. ~\ref{F_cmp}.

\subsection{NTT and INTT Dataflow}
Fig. ~\ref{F_flow} shows the CT radix-2 style NTT for the acceleration of polynomial multiplication in a ring. Each stage of NTT contains groups of butterfly operation which can be decomposed to a multiplication, an addition and a subtraction. Each stage has own grouping of points depending on the index.

  \begin{figure*}[t]
      \centering
      \includegraphics[width=0.95\textwidth]{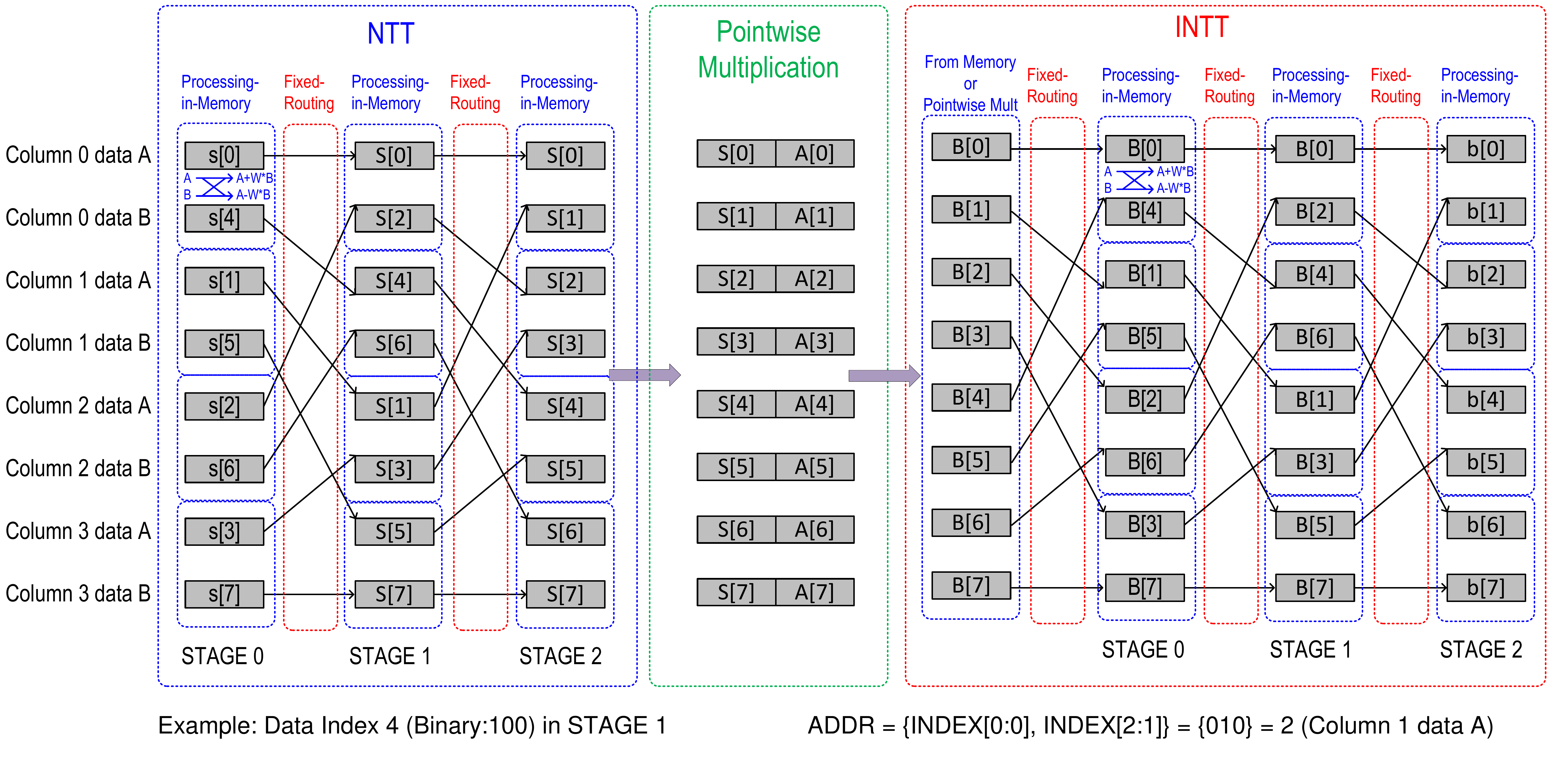}
      \vspace{-2.5ex}
      \caption{MeNTT dataflow (for computing b = INTT(NTT(A)·NTT(S)) = a*b) with a novel mapping strategy.}
      \label{F_mappingflow}
   \end{figure*}

\subsubsection{Intra-stage flow}
In a single stage, the data arrangement is shown in Fig.~\ref{F_data_arr}. The N-bit operands A and B are stored sequentially in same column followed by the twiddle factor W. The results of addition, subtraction and multiplication are computed and stored in the scratchpad area. Therefore the temporary results for an NTT stage can be computed in sequential order and updated into the operand area for follow-up computations. All the operations addition, subtraction, copying, inverting and multiplication required for a stage can be completed inside a single column in a bit-serial manner. As the algorithm for modular multiplication requires extra bits for each column, the size of scratchpad is set to 2*N+2. This space can be fully utilized to store the final outputs of a stage i..e, A+W*B and A-W*B.

\subsubsection{Inter-stage data movement}
Once the computation of a single stage is done, the data needs to be read out and written into the operands area for next stage. Traditional in-place NTT has complicated data routing, varying from stage to stage. This makes the data movement different in different stages of NTT and INTT. A configurable crossbar routing is required to enable various address matching, adding to the area and energy overheads. In~\sys \ we take advantage of unconventional data arrangement in SRAM for routing. Instead of reading column-wise, we read in traditional row-wise fashion i.e, one bit of each output. We switch these output bits and write back to different addresses. 
We propose a mapping strategy that makes this data movement between stages constant.
 Fig.~\ref{F_mappingflow} shows a complete data movement flow in a polynomial multiplication including a stage to stage routing in NTT, point-wise multiplication and INTT. The data is stored in a specific column with address. The physical address is a function of the current stage ($s$) and its original index, as described below:
 \begin{equation}
  \centering addr=\{index[log_{2}n-s-1:0],index[msb:log_{2}n-s]\}
  \label{eq:AddressMapping}
 \end{equation}
 
 For example, data S[4] in Fig.~\ref{F_mappingflow} has index `100'. In stage 1 of NTT, the actual address is \{`0',`10'\} according to the mapping, which leads to address of 2, translating to Column-1, data A. The key observation for this approach is that the actual physical output-input address mappings are the same for each stage. In the example of 8-point NTT, the output of address 4 (Column 2 data A) always goes to the input address 1 (Column 0 data B) in every stage. Therefore the crossbar connections can be reduced to simple switches. This schedule make it possible to construct a single-bank processing-in-memory block to compute the NTT process iteratively. ~\sys \ takes 4*N cycles for routing between stages, where N is bit width of the operand. In conventional digital architectures which use traditional SRAM and read row-wise, these cycles depend linearly on higher valued polynomial order.

 INTT follows a similar mapping strategy and shares the same physical routing. The point-wise multiplication between NTT and INTT can be completed in place with precomputed NTT result of another polynomial. In such a scheme, the whole polynomial multiplication operation is performed in a single SRAM bank with bit-wise modular arithmetic and constant stage to stage routing. The \sys \ therefore utilizes memory reuse for optimal area efficiency and memory bandwidth, making it suitable for resource-constraint applications.
 
\section{Evaluation and Discussion}

\begin{figure}[t!]
     \centering
     \subfloat[][energy]{\includegraphics[trim={0 0.1cm 0 0},clip, width=0.48\columnwidth]{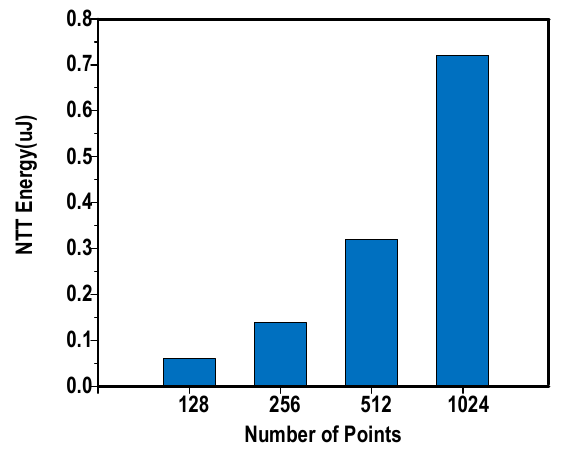}
     \label{F7_1}}
     \hfill
     \subfloat[][clock cycles]{\includegraphics[trim={0 0.1cm 0 0},clip, width=0.48\columnwidth]{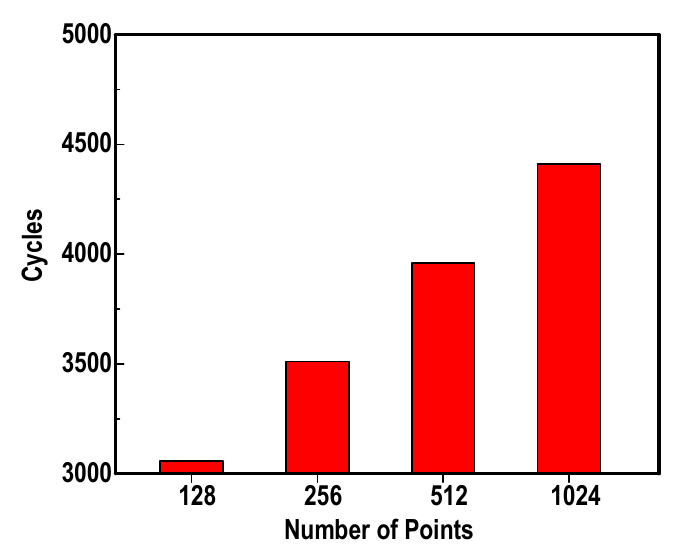}
     \label{F7_2}}
        \caption{MeNTT energy and clock cycles for different number of points at N=14b.}
        \label{fig:F7}
\end{figure}

\begin{figure}[t!]
     \centering
     \subfloat[][energy]{\includegraphics[trim={0 0.1cm 0 0},clip,width=0.48\columnwidth]{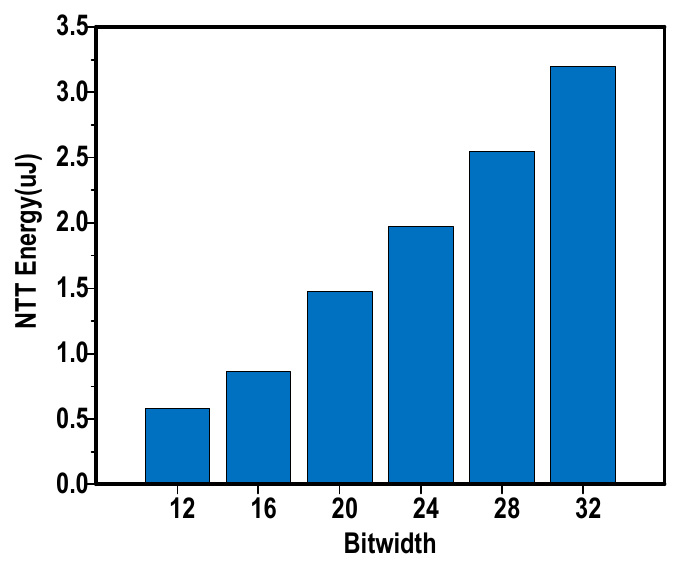}
     \label{F8_1}}
     \hfill
     \subfloat[][clock cycles]{\includegraphics[trim={0 0.1cm 0 0},clip,width=0.48\columnwidth]{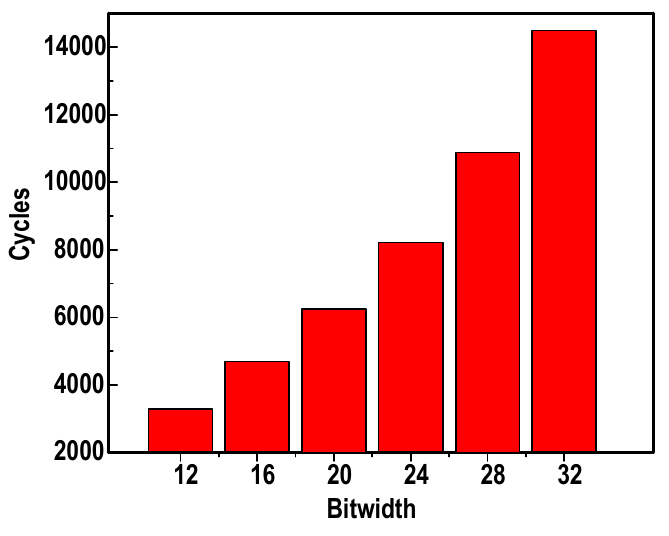}
     \label{F8_2}}
        \caption{MeNTT energy and clock cycles for different bit widths at a polynomial order of 1024.}
        \label{fig:F8}
\end{figure}

\begin{figure}[t!]
     \centering
     \subfloat[][area]{\includegraphics[trim={0.1cm 0 0.4cm 0.2cm},clip,width=0.48\columnwidth]{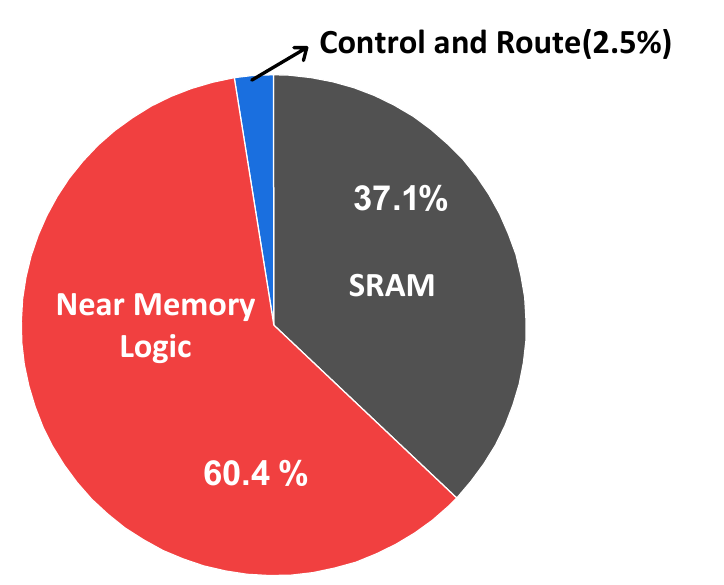}
     \label{F13_1}}
     \hfill
     \subfloat[][energy]{\includegraphics[trim={0.1cm 0 0.2cm 0.2cm},clip,width=0.46\columnwidth]{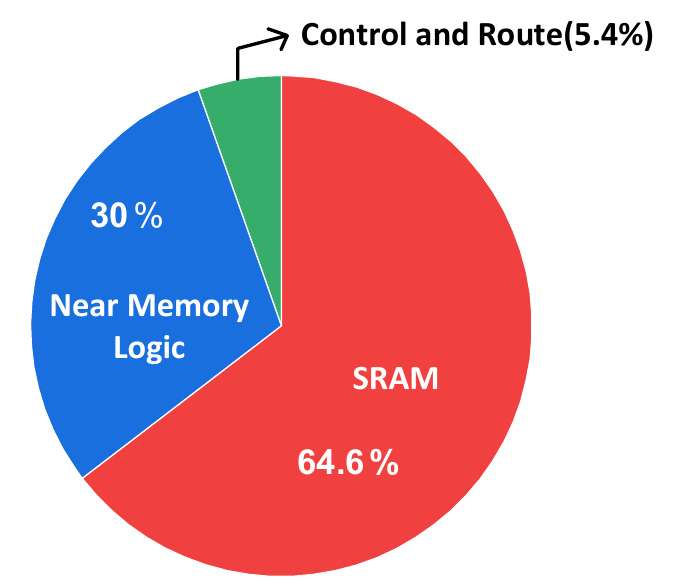}
     \label{F13_2}}
        \caption{MeNTT area and energy values breakdown for different modules at n=1024,N=32b.}
        \label{F_breakdown}
\end{figure}

We evaluated \sys \ in TSMC 65nm LP CMOS.  We implemented the digital circuits design using verilog followed by synthesis, auto-place and route (APR) by Synopsys Design Compiler. We evaluated the design using post-layout SPICE simulation of SRAM and digital circuits to estimate the overall system accurately. Different configurations of bit width and polynomial order are evaluated to compare our work with prior arts for different protocols. Evaluation results are shown in Fig. \ref{fig:F7}, Fig. \ref{fig:F8} and Fig. \ref{F_breakdown}. \sys \ is configurable for different bit width and polynomial order. Column-wise power-gating makes it possible to accommodate NTT of smaller polynomial order with energy saving. The read/write address and peripheral instructions are sent from the control block. Thanks to the routing schedule, the 6-T SRAM for PIM can process arbitrary polynomial order within the size constraint. Good scalability is maintained as SRAM can be extended to more columns in \sys. Table \ref{F_table} compares \sys \ with software, FPGA, ASIC, and previous PIM designs for NTT computation. For FPGA works, the energy and area are normalized to 65nm technology based on the gate count reported in the original paper combined with our SRAM area and energy model in 65nm node. For ASIC works, the area and energy are normalized to 65nm technology by scaling from the original technology node. For software works, the energy consumption is evaluated by the corresponding processor's reported latency, frequency, and power.

\subsection{Comparisons with software solutions}
  As discussed in previous sections, traditional software implementation has an obvious bottleneck for Ring-LWE computations with the increase of polynomial order and coefficient's bit width. The energy cost and latency are higher than hardware approaches by several orders. Under current Von-Neumann architecture, the highly parallel butterfly operations in NTT can only be executed in a serially. While modular arithmetic usually takes only one or few cycles each, the data transfer between memory and processor involves, cache and main memory read/write, which can cost 10s of cycles, making the whole process extremely expensive. From Table~\ref{F_table}, one can observe the significant energy and latency overhead in X86 based software approach. Despite the fact that a CPU can run at a higher speed than custom hardware, it is not the perfect solution for low-power and high-performance security applications. 
 Also, software solutions are not well guarded against side-channel attacks. Since all operations are executed in ALU in serial, it is relatively easy for an attacker to retrieve power and timing information leaks for a side-channel attack. Due to the low efficiency of memory access in software approaches, hardware acceleration is widely desired and adopted for FFT and NTT applications.

\subsection{Comparison with FPGA solutions}
FPGA-based hardware approaches exhibit improved performance, thanks to the customized architecture and datapath to elevate parallelism and efficiency for NTT computation. To support the modular arithmetic in NTT flow, DSP blocks \cite{poppelmann_area_2014} or custom multipliers are required to provide sufficient computing capability. Optimized datapaths are designed to accommodate the stage by stage NTT and INTT in a pipelined \cite{nejatollahi_exploring_2020} approach which achieved higher throughput with a considerable area overhead.
As shown in Table \ref{F_table}, FPGA solutions outperform the software approach in throughput and latency, but their energy and area costs are significantly higher than custom hardware. Therefore FPGA is more generally more suitable for prototyping and deployment in the cloud.

\begin{table*}[t!]
      \centering
      \caption{Comparison Table with State-of-the-Art NTT Accelerators.}
      \includegraphics[trim={0 0 0 0},clip,width=2\columnwidth]{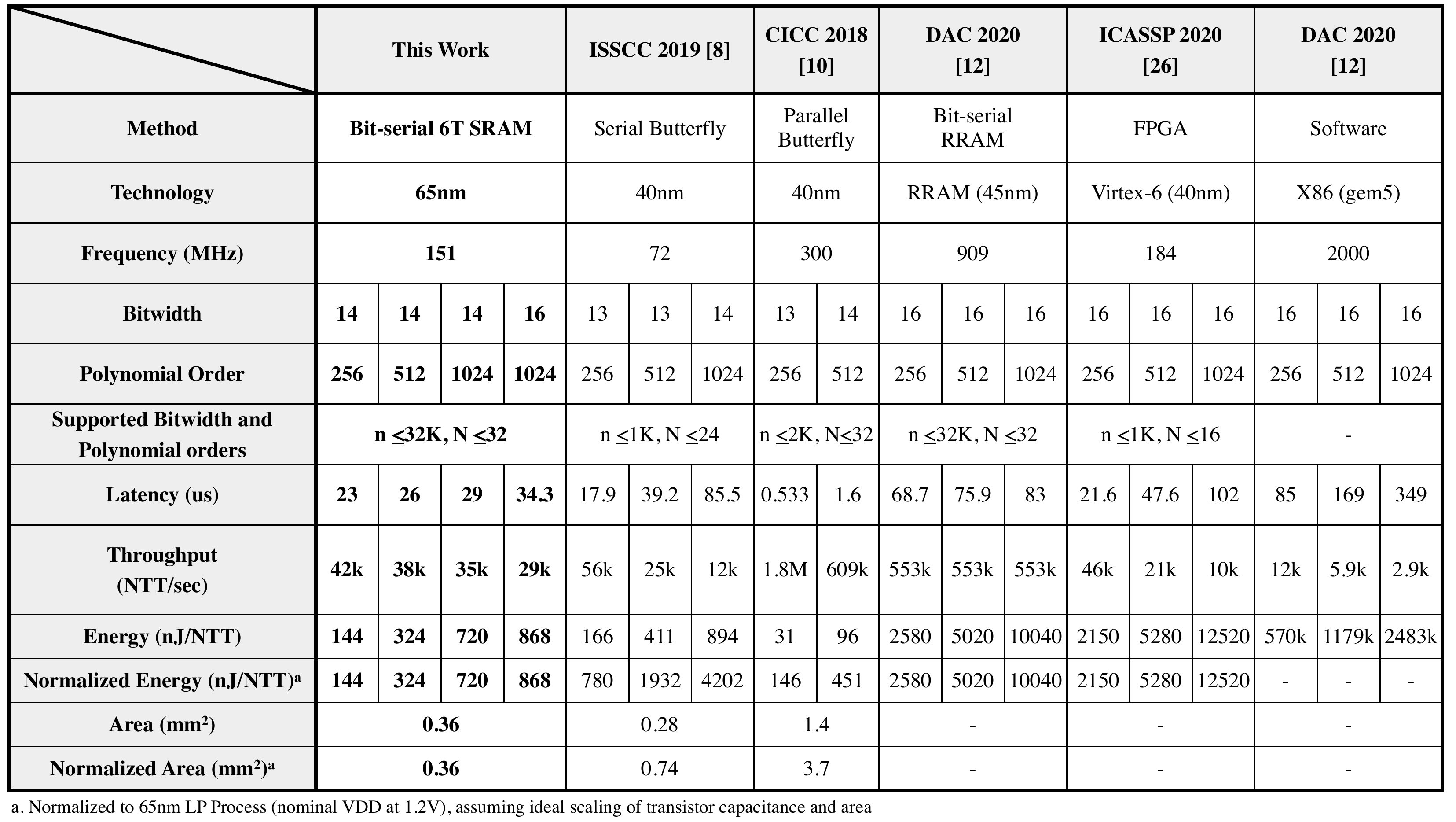}
      \label{F_table}
   \end{table*}
   
\subsection{Comparison with ASIC solutions}

The disadvantages of software and FPGA approaches inspired exploration in custom hardware solutions. ASIC implementations make use of standard SRAM or registers and compute modular arithmetic in digital circuits~\cite{banerjee_23_2019}. Although ASICs usually perform better than FPGAs in terms of speed, energy, and area, they come with much higher costs. Another major drawback for this approach is the amortized BL energy spent on reading data from SRAM. The processing speed is also limited by traditional memory bandwidth. While embedding registers for local storage in each computing unit increases throughput by increasing parallelism as in~\cite{song_leia_2018}, the area overhead and limited scaling potential are the main drawbacks. 

In terms of scalability to larger bit width and q, \sys \ executes modular arithmetic in bit-serial and word-parallel order while traditional software, FPGA, and ASIC approaches carry out butterfly unit operations word by word. Thus, \sys \ will achieve even higher throughput and energy efficiency than the other solutions, when the polynomial order gets higher.

ASIC solutions require higher design and fabrication cost, and need extra overhead to handle different schemes with different polynomial order, q and bit width. \sys \  provides higher degree of configurability by providing general modular arithmetic in a highly parallel computing approach.  The computation and data movement can be reprogrammed with minimal effort by changing WL accessing sequence and peripheral configurations. The SRAM rows and columns can be gated in different cryptography schemes for higher throughput and energy savings. \sys \ also provides higher normalized energy and area efficiency, as shown in Table \ref{F_table}, mainly benefiting from in- and near-memory computation.

\subsection{Comparisons with existing PIM solutions}
Compared to previous PIM studies that focus on leveraging parallel word-serial or bit-serial operation for general-purpose arithmetic in SRAM~\cite{recryptor} and ReRAM~\cite{nejatollahi2020cryptopim,eckert_neural_2018}, 
optimized modular arithmetic and dataflow help~\sys  \ outperforms prior PIM works in energy and area efficiency as well as latency. \sys \ masks the inherent modular reduction cycles in the modular multiplication operation itself, whereas other designs use Barret or Montgomery reduction techniques which cause an extra area and latency overheads.

\begin{figure}[t!]
     \centering
     \subfloat[][latency]{\includegraphics[trim={0 0 0 0},clip, width=0.48\columnwidth]{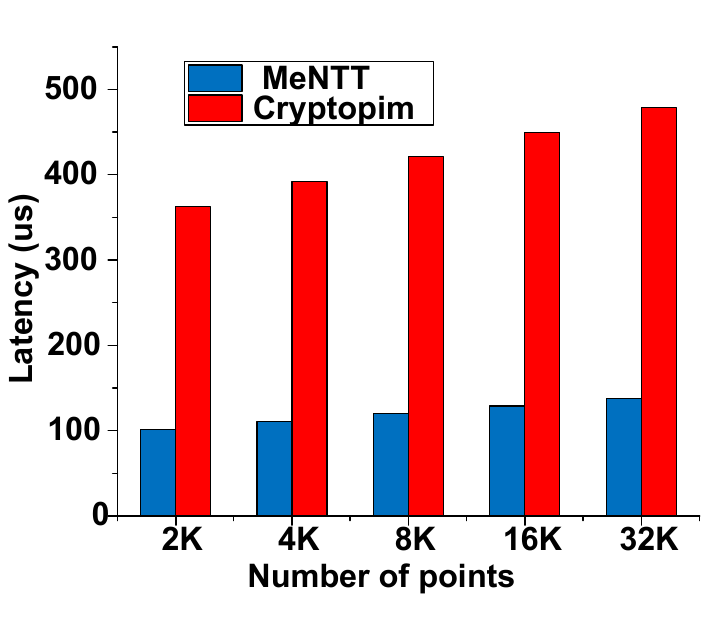}
     \label{F14_1}}
     \hfill
     \subfloat[][energy]{\includegraphics[trim={0 0 0 0},clip, width=0.48\columnwidth]{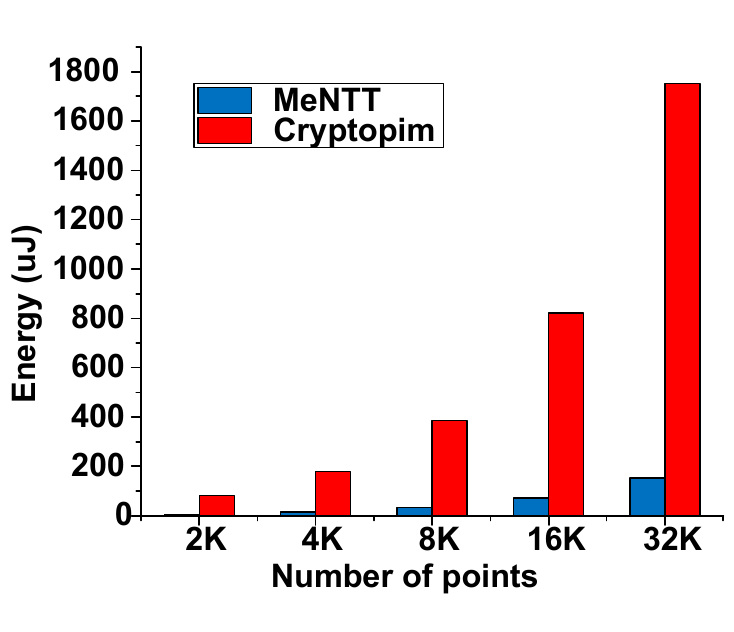}
     \label{F14_2}}
        \caption{Comparison of latency and energy values for MeNTT vs Cryptopim for different polynomial orders at 32 bit width. }
        \label{fig:F14}
\end{figure}

While ~\cite{nejatollahi2020cryptopim} achieved higher throughput by introducing multiple pipelines to break the data path into shorter pieces, our approach focuses on designing ultra-compact single SRAM bank implementation, which is more applicable to resource-constrained applications. Note that the reported latency for \cite{nejatollahi2020cryptopim} is for specially selected q with very small hamming weight. The latency varies a lot with different choices of schemes and q. Other generic choices of q will incur high latency overhead.
Last but not least, while the \sys \space design is implemented and evaluated with CMOS devices and SRAMs, the proposed modular arithmetic protocol and dataflow techniques are generic to any bit-serial PIM fabrics and can be easily adopted in PIM with emerging memories.





\section{Conclusion}
In conclusion, this paper presents \sys, a novel PIM architecture for NTT acceleration. With the proposed bit-serial modular arithmetic protocol and mapping strategy, it achieves superior efficiency and throughput with a compact footprint. A fully functional mixed-signal implementation of the system verifies its feasibility in physical design, and provides realistic estimation of its performance for comparison.


%








\bibliographystyle{IEEEtran}
\bibliography{IEEEabrv, PQC,TVLSI_Extra}
%


%



\begin{IEEEbiography}[{\includegraphics[width=1in,height=1.25in,clip,keepaspectratio]{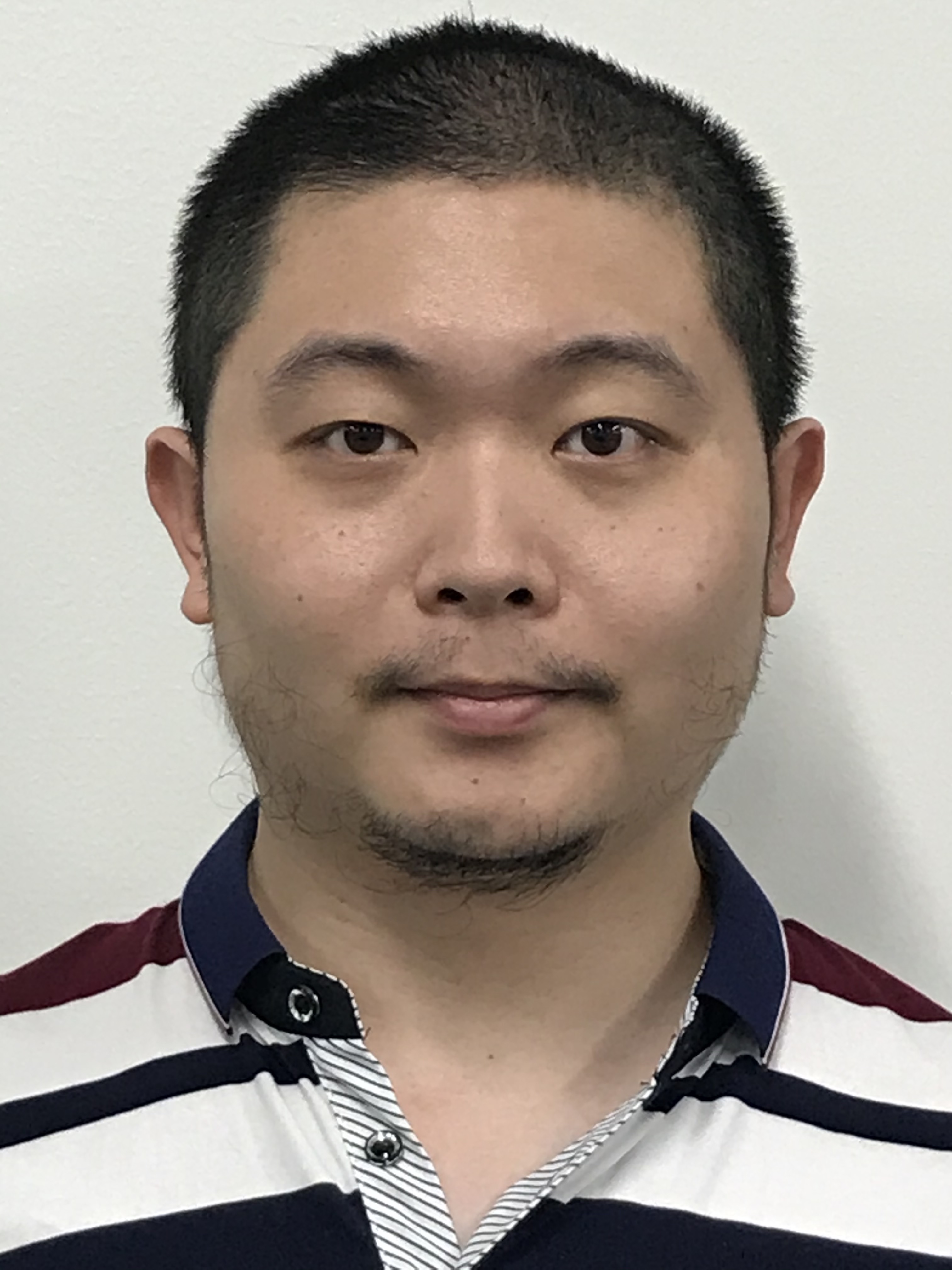}}]{Dai Li}

received the B.S. and M.S. degrees in electronic engineering from Tsinghua University, Beijing,
China, in 2010 and 2013, respectively, and the PhD degree in electrical and computer engineering from
Rice University, Houston, TX, USA, in 2021. He is now at Google.

His research interests include very large-scale integration (VLSI) circuits, hardware security, mixed-signal integrated circuits, and low-power circuits.

\end{IEEEbiography}

\vskip -2\baselineskip plus -1fil
\begin{IEEEbiography}[{\includegraphics[width=1in,height=1.25in,clip,keepaspectratio]{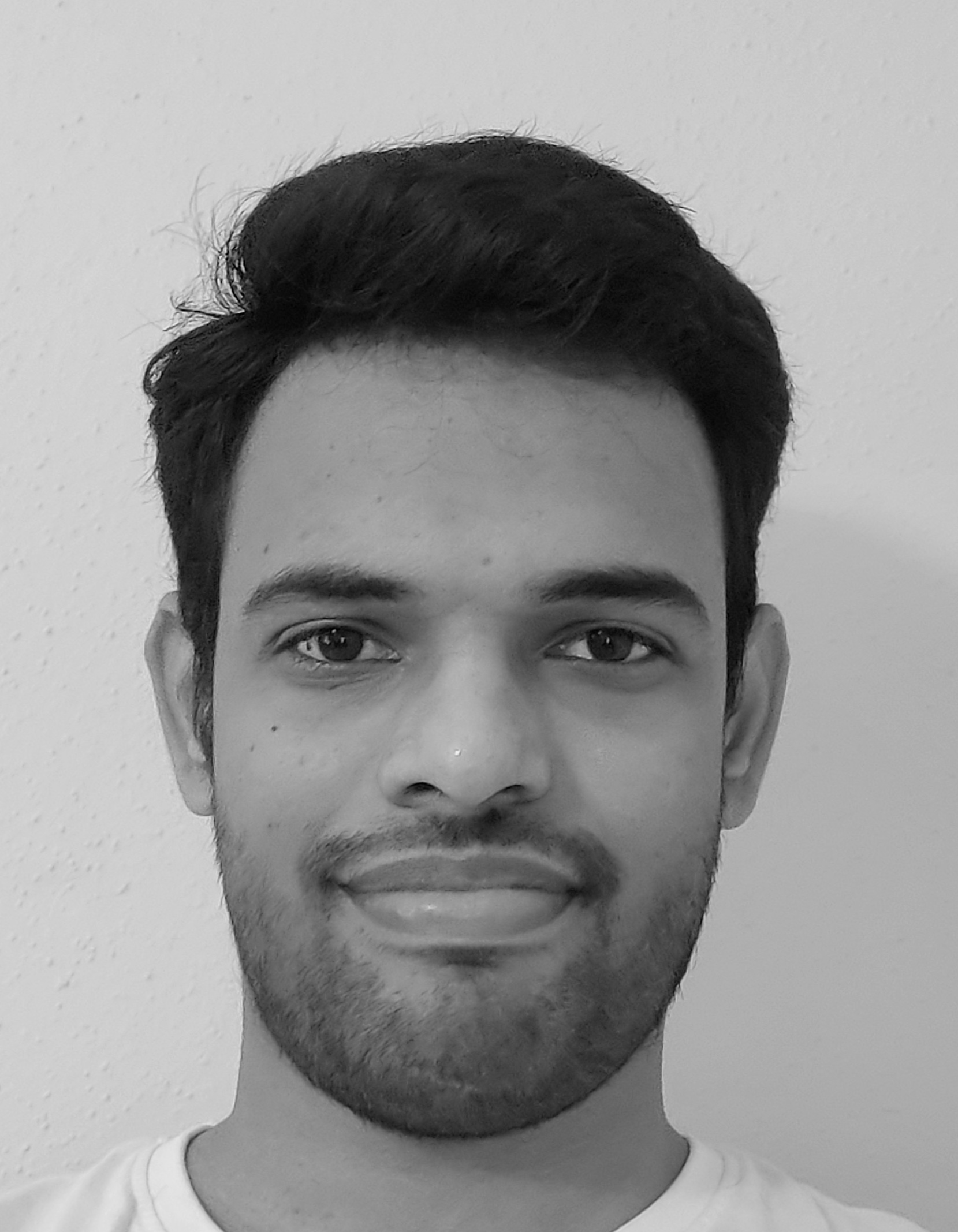}}]{Akhil Pakala}
 received his Bachelor’s and Master’s degree from the Indian Institute of Technology Madras, Chennai, India, in 2019. From 2019-2020, he has worked with Samsung Semiconductor Research and Development, India on SerDes PHY IP. He is currently pursuing his doctoral degree at Rice University, Houston, Texas.
 
 His current research interests include designing digital and mixed-signal circuits for security and machine learning applications.
\end{IEEEbiography}

\vskip -2\baselineskip plus -1fil
\begin{IEEEbiography}[{\includegraphics[width=1in,height=1.25in,clip,keepaspectratio]{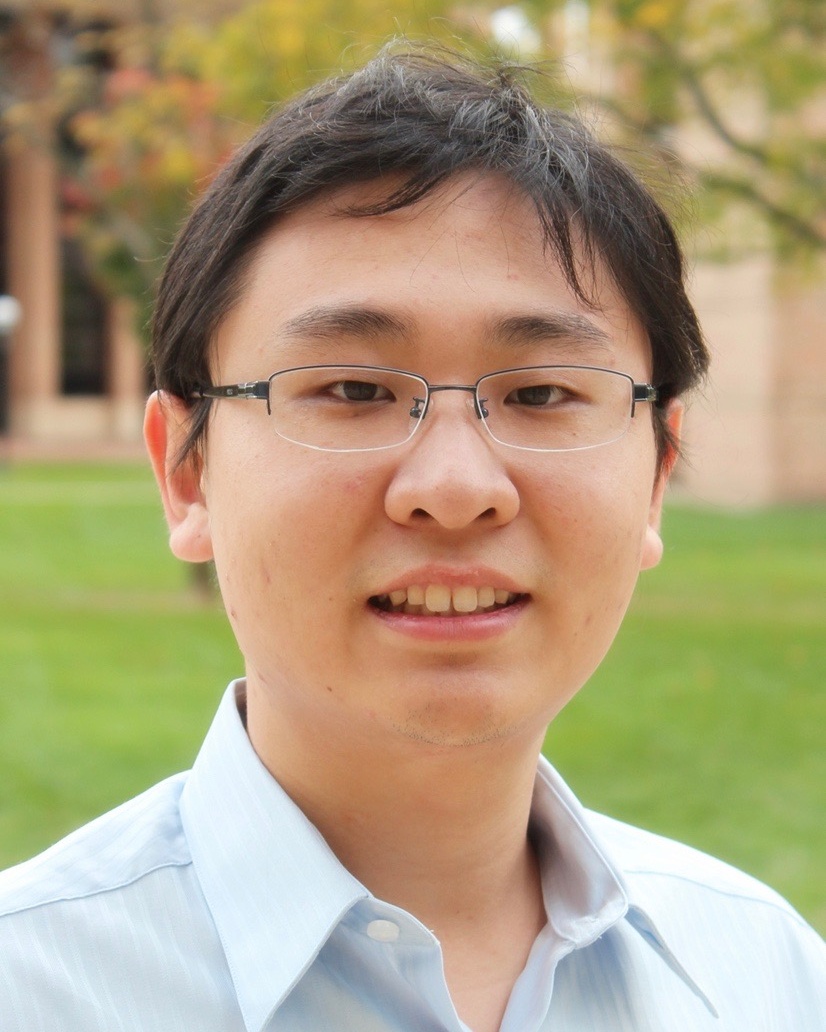}}]{Kaiyuan Yang}
(S'13-M'17) received the B.S. degree in Electronic Engineering from Tsinghua University, Beijing, China, in 2012, and the Ph.D. degree in Electrical Engineering from the University of Michigan, Ann Arbor, MI, in 2017. His Ph.D. research was recognized with the 2016-2017 IEEE Solid-State Circuits Society (SSCS) Predoctoral Achievement Award. 

He is an Assistant Professor of Electrical and Computer Engineering at Rice University, Houston, TX. 
His research interests include low-power digital/analog/mixed-signal integrated circuit and system design for secure and intelligent micro-systems, hardware security, and bioelectronic applications.
Dr. Yang received a number of best paper awards from major conferences in various fields, including the Best Paper Award at the 2021 IEEE Custom Integrated Circuit Conference (CICC), Distinguished Paper Award at the 2016 IEEE International Symposium on Security and Privacy (Oakland), Best Student Paper Award (1st place) at the 2015 IEEE International Symposium on Circuits and Systems (ISCAS), the Best Student Paper Award finalist at the 2019 IEEE Custom Integrated Circuit Conference (CICC), and the 2016 Pwnie Most Innovative Research Award Finalist. His research was selected as “Research Highlight” of Communication of ACM in 2017.

\end{IEEEbiography}





\end{document}